\documentstyle[floats,twocolumn,prl,aps,psfig]{revtex} 

\psfigurepath{. :.}
\psrotatefirst

\tighten

\begin{document}
\draft
\preprint{DRAFT}

\wideabs{
\title{Capillary Waves at Liquid/Vapor Interfaces:
A Molecular Dynamics Simulation}

\date{\today}

\author{
Scott W. Sides,  $^{\dagger}$
Gary S. Grest,   $^{\dagger}$
and Martin-D. Lacasse$^{\ddagger}$
}

\address{
$^{\dagger}$Sandia National Laboratories,
              Albuquerque, New Mexico 87185-1411\\
$^{\ddagger}$Corporate Research Science Laboratories,
              Exxon Research and Engineering Company,
              Annandale, New Jersey 08801
}

\maketitle
\begin{abstract}
Evidence for capillary waves at a liquid/vapor interface
are presented from extensive molecular dynamics simulations of
a system containing up to $1.24$ million Lennard-Jones particles.
Careful measurements show that the total interfacial width
depends logarithmically on $L_\parallel$,
the length of the simulation cell parallel to the interface,
as predicted theoretically.
The strength of the divergence of the interfacial
width on $L_\parallel$ depends inversely on the
surface tension $\gamma$. This allows us to measure
$\gamma$ two ways since $\gamma$
can also be obtained
from the difference in the pressure
parallel and perpendicular to the interface.
These two independent measures of $\gamma$ agree provided that
the interfacial order parameter profile is fit to an error function and
not a hyperbolic tangent, as often assumed.
We explore why
these two common fitting functions give different results for
$\gamma$.
\end{abstract}
}

\narrowtext


An interface is the physical boundary between
two distinct thermodynamic phases, i.e.
a region characterized by a local gradient
of the order-parameter which mean value changes
from one phase to the other.
Examples include domain boundaries in ferromagnetic materials,
the interface between two immiscible liquids, or between a liquid and
its own vapor below the critical temperature $T_c$. 
This last case has been well studied, both
theoretically and experimentally.
For simple fluids interacting via van der Waals forces, 
the mean local density
changes monotonically \cite{evans94} across the interface
from its bulk liquid value to that of the vapor.
In other systems, such as alkali metals for example \cite{rice96,rice98},
the profile across the interface is often more complex, with oscillations
in the local density superimposed on the decaying density profile.

For simple fluids, 
thermodynamic considerations alone would predict that
the interfacial width $w$, depends
only on temperature and on the interaction energies within each phase and
across the interface.
However, the presence of the interface breaks the translational invariance of
the system, inducing Goldstone fluctuations or ``capillary waves''
at an interface \cite{buff65,widom82_book}.
For two-dimensional interfaces,
these non-critical fluctuations give rise to a logarithmic increase
in the interfacial width $w$ with increasing $L_\parallel$,
the length of the interface.
Evidence for capillary waves has been
found experimentally from X-ray scattering \cite{sany91,tids91,tola98}
on liquid/vapor interfaces and neutron reflectivity
\cite{shull93,sfer97,stamm99} on polymer/polymer interfaces.
Moreover, nuclear reaction analysis (NRA) depth profiling
\cite{kerl96} has been used to directly investigate the
film thickness dependence on the interface
width between two polymer films and is in qualitative agreement with
capillary-wave predictions.
Capillary waves have also also been observed in
computer simulations for polymer/polymer interfaces
\cite{kerl96,wern97,wern99,wern99_2,lacasse98}.
Most previous simulations \cite{nijm88,adams91}
of the liquid/vapor interface in three dimensions
did not investigate the dependence of $w$ on the size of the interface.
One recent simulation study \cite{heer98} of a thin
polymer-film system gave some
evidence for capillary waves, but the longitudinal size of the interface
was very small.

The purpose of this paper is
to present computer simulation results of interfaces in a liquid/vapor
system.
To our knowledge, these simulations are the most extensive studies of
the interface fluctuations due to capillary waves.
In particular, we obtain the surface tension $\gamma$ two different ways:
from the dependence of $w$ on $L_\parallel$ $(\gamma_w)$,
and from the difference in pressure parallel $p_\parallel$ 
and perpendicular $p_\perp$ to the interface $(\gamma_p)$.
We find the surprising result, that $\gamma_w$ depends on the
functional form chosen to fit the order parameter
profile through the interface.
In particular, fitting the profile to an error function gives results
for $\gamma_w$ which are in excellent agreement with $\gamma_p$.
However, fitting our data to $\tanh(2z/w)$, a functional
form derived from mean-field arguments \cite{widom82_book},
gives results
for $\gamma_w$ which are systematically 15\% smaller than $\gamma_p$.
Since the $\tanh$ function is often used to fit interfacial
profiles at the liquid/vapor interface \cite{wern97,heer98},
this difference is important to understand.

For this study we perform continuous-space, molecular dynamics simulations on
a system of particles interacting through a standard
(12-6) Lennard-Jones potential.
The potential between particles $i$ and $j$ takes the form
 \begin{equation}
 \label{eq_ljpotential}
 \nonumber
  U(r_{ij})
   =
   \left \{ \begin{array}{ll}
    4 \epsilon \left [ \left (\frac{\sigma}{r_{ij}} \right )^{12} -
                       \left (\frac{\sigma}{r_{ij}} \right )^{6}
               \right ]  & r_{ij} < r_{c}, \\
    0                    & r_{ij} > r_{c},
   \end{array} \right.
 \end{equation}
where
$r_{ij}$ is the distance between particles $i$ and $j$, and
$\epsilon$ and $\sigma$ set the energy and length scales
of the potential respectively.
Here we take a cut-off of $r_{c}$=$2.5 \sigma$.
Increasing $r_{c}$ merely shifts $T_c$ to higher values, which should have
little effect on the capillary-wave properties while increasing computation
time significantly.
The trajectories of the $N$ particles of mass $m$,
are obtained by stepwise integration of 
Newton's equations of motion (EOM)
 \begin{equation}
 \label{eq_eom}
 \nonumber
  m \frac{d^2 r}{dt^2}
  =
  - \nabla U(r) - m \Gamma \frac{dr}{dt} + W(t) \ .
 \end{equation}
In addition to the force derived from the LJ potential, the EOM contains
a velocity-dependent damping term and a noise term representing
a viscous drag force and a weak stochastic force, respectively.
The noise term $W(t)$ is taken from a uniform distribution, which
mean value is set from the
temperature $T$ and the damping coefficient $\Gamma$
through the fluctuation-dissipation theorem \cite{doi_book}.
The combination of the viscous damping and stochastic force terms in the
EOM effectively couples the system to a heat bath.
Our simulations are performed in the canonical ensemble
with fixed particle number and volume (constant-NVT).
The EOM for each particle is integrated with the
velocity-Verlet \cite{tild87} algorithm with a time step 
$\Delta t=0.006 \tau$, where
$\tau$=$\sigma (m/\epsilon)^{1/2}$ fixes the time scale. We set 
$\Gamma$=$0.5 \tau^{-1}$.
All results presented here are measured in reduced units, as derived
from the fundamental scales fixed by $\sigma$, $\epsilon$, $m$, and
the Boltzmann constant $k_B$.
To reduce computation time we use a combination of the Verlet
and linked-cell list algorithms \cite{tild87}.

\begin{table}
 \begin{center}
 \begin{tabular}{c | c c c c}
  $T$ $[\epsilon/k_{\rm B}]$ &
  $N$                        & 
  $L  [\sigma]$                 &
  $L_{\perp} [\sigma]$         &
  time $[\tau]$  \\
  \hline \hline
  0.8
  & $7200$        & $12.8$    & $127.0$    & $6000$             \\
  & $24000$       & $24.7$    & $127.0$    & $6000$             \\
  & $69360$       & $42.0$    & $127.0$    & $6000$             \\
  & $154400$      & $54.5$    & $195.6$    & $5000$             \\
  & $506880$      & $94.0$    & $125.6$    & $2800$             \\
  \hline
  0.9
  & $14400$       & $15.1$    & $216.1$    & $11000 $           \\
  & $40000$       & $25.2$    & $216.1$    & $10000 $           \\
  & $115660$      & $42.9$    & $216.1$    & $7800  $           \\
  & $154400$      & $54.5$    & $195.6$    & $5800  $           \\
  & $506880$      & $94.0$    & $144.4$    & $4200  $           \\
  & $1240000$     & $134.6$   & $164.4$    & $4200  $           \\
  \hline
  1.0
  & $14400$       & $13.3$    & $264.2$    & $16500 $           \\
  & $48000$       & $25.7$    & $264.2$    & $14400 $           \\
  & $138720$      & $43.7$    & $264.2$    & $10500 $           \\
  & $170000$      & $54.5$    & $195.6$    & $13500 $           \\
  & $590000$      & $94.0$    & $293.9$    & $4900  $           \\
 \end{tabular}
 \vspace{0.2in}
 \caption[Simulation parameters]
 { \label{table_simul_para}
 Values used for the parameters of simulations: temperature $T$,
 number of particles $N$, $L=L_\parallel$, $L_\perp$,
 and duration of run.
 }
 \end{center}
\end{table}

\begin{figure}[tb]
\centerline{\psfig{file=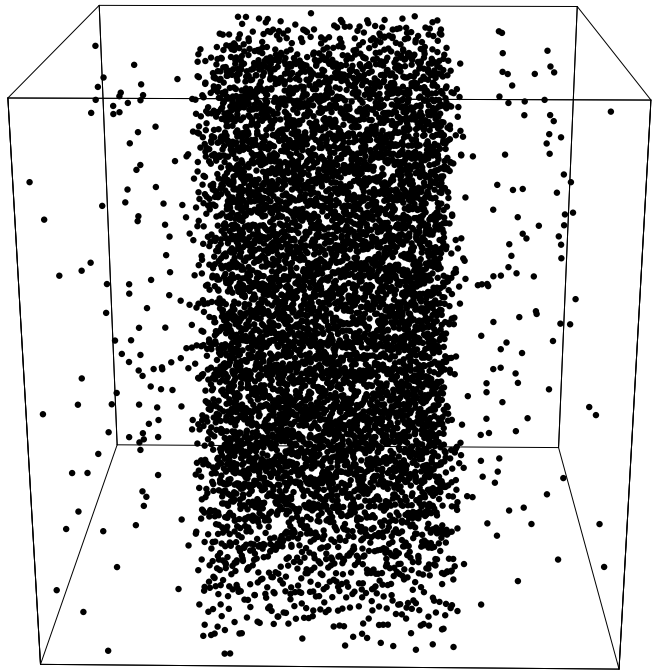,angle=270,width=\hsize,%
bbllx=15pt,bblly=105pt,bburx=660pt,bbury=710pt}}
\caption
{\label{fig_config}
Typical configuration of an equilibrated liquid/vapor interface
at $T$=$0.8 \ \epsilon/k_{\rm B}$. Length of square
cross section holding the interface is $L$=$12.8 \sigma$.
}
\end{figure}

Periodic boundary conditions are used in all three space dimensions,
thus forcing the creation of (at least)
two interfaces in a two-phase system.
The system sizes, temperatures, particle numbers and equilibration
times of our simulations are listed in Table~\ref{table_simul_para}.
In Table~\ref{table_simul_para} and throughout this paper, $L$ refers
to the dimensions of the square cross section
parallel to the interface, which lies in the
$xy$ plane. Thus, $L_{x}$=$L_{y}$=$L_{\parallel}$=$L$.
$L_{\perp}$ refers to the dimension of the box perpendicular to the
plane of the interface.
Simulations are performed for $L$ ranging from $12.8 \sigma$ to 
$134.6 \sigma$.
The largest system we could run contains $1.24$ million particles.
After that, computation becomes prohibitively slow
due to the large number of particles.
At the other end of our size range, systems with $L<12 \sigma$
demonstrate non-negligible finite-size effects.
Figure \ref{fig_config} shows a typical configuration of an
equilibrated system of 
$L$=$12.8 \sigma$ at $T$=$0.8 \epsilon/k_{\rm B}$.

\begin{table}[bt]
 \begin{center}
 \begin{tabular}{c || c c c c }
  $T$ $(\epsilon/k_{\rm B})$ &
  $\rho_V$                   &
  $\rho_L$                   &
  $\gamma_{w_{e}}$           &
  $\gamma_p$                 \\
  \hline \hline
  0.8
  & $0.020(1)$       & $0.730(1)$   & $0.37(3)$   & $0.39(1)$          \\
  \hline
  0.9
  & $0.045(1)$       & $0.663(1)$   & $0.22(1)$   & $0.22(1)$          \\
  \hline
  0.95
  & $0.066(1)$       & $0.623(1)$   &  ----       & $0.15(1)$          \\
  \hline
  1.0
  & $0.098(2)$       & $0.571(1)$   & $0.097(2)$  & $0.08(1)$          \\
 \end{tabular}
 \vspace{0.2in}
 \caption[Simulations results]
 {\label{table_results}
 Calculated values of the bulk densities and surface tensions
 for different simulation temperatures.
 }
 \end{center}
\end{table}

Initial systems were built as follows:
for each system size and temperature, we construct a
slab of the liquid phase
and center it in the middle of the simulation cell with the interface
perpendicular to the $z$ direction.
$L_{\perp}$ is set such that it is at least twice the length of
the liquid slab, allowing sufficient space for the bulk liquid and
vapor densities to achieve constant values.
Since the phase coexistence diagram is well known for this system
\cite{nijm88,adams91,smit92}, we adjusted the density of the liquid
and vapor regions to be close to their reported values for each
temperature $T$.

\begin{figure}[thb]
\centerline{\psfig{file=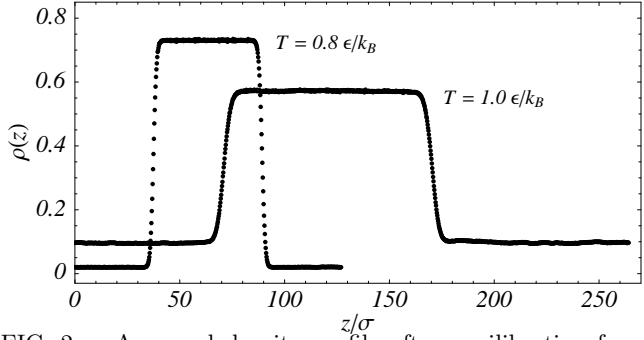,angle=270,width=\hsize,%
bbllx=130pt,bblly=0pt,bburx=520pt,bbury=760pt}}
\caption
{\label{fig_denspro}
Averaged density profile after equilibration for
$T$=$0.8 $ and $1.0 \ \epsilon/k_{\rm B}$
for  $L$=$41.9 \sigma$.
}
\end{figure}

\begin{figure}[thb]
\centerline{\psfig{file=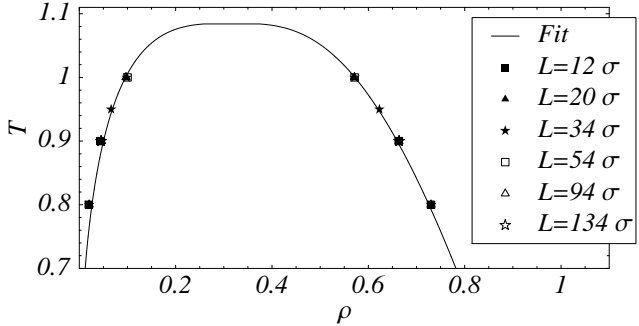,angle=270,width=\hsize,%
bbllx=150pt,bblly=60pt,bburx=505pt,bbury=760pt}}
\caption
{\label{fig_coex}
Coexistence curve: bulk density values are
obtained from tail values of interfacial
liquid/vapor density profiles.
Curve is a best fit using the following expressions
\protect\cite{adams91},
$0.5(\rho_L+\rho_V)\sigma^3$=$0.544-0.210 k_B T/\epsilon$
and
$(\rho_L-\rho_V)\sigma^3$=$A(1-(T/T_c))^{0.318}$.
The best fit parameters are
$A$=$1.07$ and $T_{c}$=$1.085$.
}
\end{figure}

After the system has equilibrated the density profile is
measured, i.e. the $xy$-cross-section averaged
number density $\rho(z)$ as a function of $z$.
Over the course of a simulation for a given $T$ and $L$,
$\rho(z)$ is measured every $400$ time steps.
Once the interfaces have equilibrated, the density
profiles are averaged over $10^5$ -- $2\times 10^5\Delta t$.
Great care must be exercised in the averaging procedure.
For each profile, the position of the interface is located to insure
that the averaging does not artificially broaden the interface width
due to drift in the interface positions.
Figure \ref{fig_denspro} shows an example of an equilibrated, averaged
density profile for $T=0.8$ and $1.0\epsilon/k_{\rm B}$ for 
$L$=$41.9 \sigma$.

Bulk values for the density are extracted from the tail values
(obtained through a fit described below)
of the interfacial density profiles.
Our final equilibrated values for the bulk liquid and vapor
densities are listed in Table~\ref{table_results}, and agree
very well with values reported in the literature.
The derived coexistence curve, along with a fit to
an expression suggested
in Ref.~\cite{adams91} are shown in Fig.~\ref{fig_coex}.
The very good agreement of the bulk values suggests that
our systems are well equilibrated that our measurement
procedures are sound.
Since the simulations are started near their respective
liquid and vapor values,
the bulk density values shown in Fig.~\ref{fig_coex} all attained
equilibrium values quickly.
However, the interface structure did not equilibrate until the simulations
had been run for the much longer times
shown in Table \ref{table_simul_para}.

An important quantity characterizing the interface is the width.
The intrinsic width of an interface is due to the intermixing of the
two phases, which always occurs to a certain degree
at finite, subcritical temperatures.
In addition to
this mixing, capillary-wave theory \cite{buff65}
predicts that thermal fluctuations
of the location of the interface will contribute
to the total, cross-section averaged, measured width.
This broadening depends primarily on the surface tension,
the temperature, and the cross-sectional size of the interface,
and the spatial dimension.
As an example, capillary-wave theory states
that any two-dimensional
crystal is unstable against thermal fluctuations \cite{huan87}.

Fluctuations in $\zeta(x,y)$,
the mean location of the interface in the $z$ direction,
induces fluctuations in the total area of the interface
and can be easily determined by expanding the
shape of the interface to first order.
This approximation is accurate
provided the interface is smooth, with no overhangs.
The free energy of the interface is the product of
its surface area and an interfacial energy density $\gamma$,
which is assumed to be independent of local curvature.
Fluctuations due to capillary waves have an energy cost due to the
increase in the surface area of the interface.
The resulting interfacial Hamiltonian can be expressed as
the product of surface tension
times the increase in interfacial area
 \begin{eqnarray}
 {\mathcal{H}} \{\zeta \}
  & = &
  \gamma \int dx \ dy \ \left [ \sqrt{1+\frac{\partial \zeta}{\partial x}}
                                \sqrt{1+\frac{\partial \zeta}{\partial y}} - 1
                        \right ], \\
  & \approx &
  \frac{\gamma}{2} \int dx \ dy \ |\nabla \zeta(x,y)|^{2} \ .
 \end{eqnarray}
The capillary-wave spectrum can be calculated by substituting the Fourier
transform of $\zeta$ giving
 \begin{equation}
 {\mathcal{H}} \{\zeta \}
  \approx 
  \frac{\gamma}{2} \int d\vec{q} \ q^2 |\tilde{\zeta}(\vec{q})|^{2}
 \end{equation}
where $\vec{q}$ represents a two-dimensional vector in reciprocal space, and
${\mathcal{F}} \left [ \zeta(x,y) \right ]$=$\tilde{\zeta}(\vec{q})$
is the Fourier transform of $\zeta(x,y)$.
The equipartition theorem dictates the mean-square amplitude
for each interfacial excitation mode,
 \begin{eqnarray}
  \left < |\tilde{\zeta}(\vec{q})|^{2} \right >
     & = & \frac{k_{\rm B} T}{4 \pi^2 \gamma q^2},
\end{eqnarray}
and summing over all allowed modes, one gets
\begin{eqnarray}
  \left < |\tilde{\zeta}|^{2} \right >
     & = & \frac{k_{\rm B} T}{4 \pi^2 \gamma}
           \int^{q_{max}}_{q_{min}} \frac{d\vec{q}}{q^2}, \\
     & = & \frac{k_{\rm B} T}{2 \pi \gamma}
           \ln \left( \frac{L}{B_{o}} \right ),
 \end{eqnarray}
where $q_{\rm min}$=$2\pi/L$ and $q_{\rm max}$=$2\pi/B_{o}$
Note that both lower and upper cut-offs are required
to prevent the value of
the integral from diverging.
The long-wavelength cutoff $q_{\rm min}$, is determined by
$L_{\parallel}$ \cite{gravity}.
The interpretation of the short-wavelength cutoff,
$q_{\rm max}$, is not as clear.
Werner et al.~\cite{wern99,wern99_2}
have studied the dependence of $B_{o}$ for polymer/polymer interfaces
and suggest $B_{o}$ scales inversely with the molecular weight.
However, the exact nature of this short-wavelength cut-off
remains an open question.

In both simulations and experiments, the quantity measured
is the {\it total} interfacial width, which includes contributions
from the intrinsic width and the broadening due to capillary-wave
fluctuations.
The two effects can be distinguished if
one assumes that capillary-wave fluctuations are decoupled
from the intrinsic profile.
Therefore, the total interface profile $\Psi(z)$ may be expressed as a
convolution of the intrinsic interface profile $\psi(z)$
and the effect due to capillary waves \cite{lacasse98},
 \begin{eqnarray}
 \label{eq_psi}
  \Psi(z)
  & = &
  \int^{\infty}_{-\infty} \psi(z-z_o) \ P(z_o)\ dz_o\ . 
 \end{eqnarray}
Here, $P(z_o)$ is the probability of finding the interface at $z_o$,
i.e.,
\begin{eqnarray}
P(z_o)
  & = &
\frac{1}{L_xL_y}\int_0^{L_x}\int_0^{L_y} dx\ dy\ \delta(\zeta(x,y) - z_o).
\end{eqnarray}
The interfacial order parameter
profile $\Psi(z)$ is related to the cross-section averaged density
profile $\rho(z)$ by the function
 \begin{equation}
 \label{eq_scale}
  \Psi(z) = 
  \frac{2}{\rho_L-\rho_V}
  \left  [
  \rho(z) - \frac{\rho_L+\rho_V}{2}
  \right ],
 \end{equation}
which scales the density profile so that $\Psi(z)$ varies between
$-1$ and $1$.
The variance of the derivative of the total profile
$d \Psi(z) / dz \equiv \Psi'$ can be used as a measure of the width
of the interface.
The variance of a distribution $f$ is given by
 \begin{eqnarray}
 \label{eq_v[f]}
  v[f] = \frac{\int^{\infty}_{-\infty} z^2 f(z) \ dz}
              {\int^{\infty}_{-\infty}     f(z) \ dz}
       = \frac{-\frac{d^2}{dq^2} \tilde{f}(q)|_{q=0}}
              {\tilde{f}(0)},
 \end{eqnarray}
where $\tilde{f}(q)$ is the Fourier transform of $f(z)$.
Making use of the convolution theorem and Eqs.~(\ref{eq_psi}) and (\ref{eq_v[f]})
it can be shown that \cite{lacasse98}
 \begin{eqnarray}
  \nonumber
  v[\Psi'] & = & v[\psi'] + v[P], \\
  \label{eq_delta2}
  \Delta^2 & = & \Delta_{o}^{2}+
                 \frac{k_{\rm B} T}{2 \pi \gamma}
                 \ln \left( \frac{L}{B_{o}} \right ).
 \end{eqnarray}
The squared widths of the total and intrinsic interfacial profiles
have been defined as $\Delta^2 \equiv v[\Psi']$ and
$\Delta_{o}^2 \equiv v[\psi']$, respectively.
Note that the average squared fluctuations of the
interface about its mean location
in the $z$ direction can be directly identified as
$\left < |\zeta|^{2} \right >$=$v[P]$.
Thus, our choice of measure for the interfacial width
clearly shows that the total interfacial width
can be written as the sum of an intrinsic part
and a contribution due to capillary-wave fluctuations.

In order to verify this prediction,
we performed several simulations on different
system sizes.
Traditionally, the order parameter interfacial profile has been fit
with $f(z) = \tanh(2z/w_t)$ or an error function ${\rm erf}(\sqrt{\pi}z/w_e)$.
Using our data
we can test these two fitting functions and the resulting
predictions for $\gamma$.
Another reason for fitting our
results for $\Psi(z)$ to one of these two functions is that
we found we can determine a value for $\Delta^2$
more accurately than by extracting it directly from the data;
once the fitting parameters of $f(z)$ have been determined, 
$v[f]$ can be easily calculated.
The two different fitting functions we tested are
 \begin{eqnarray}
 \nonumber
  f_e(z,w_e) & = & {\rm erf} \left ( \frac{\sqrt{\pi} z}{w_e} \right ), \\
  f_t(z,w_t) & = & \tanh \left ( \frac{2 z}{w_t} \right ) \ .
 \end{eqnarray}
For these two functions, the variance
of each in terms of their associated widths $w_x$'s are 
 \begin{equation}
 \nonumber
  \Delta^2
   \rightarrow
   \left \{ \begin{array}{ll}
  v[f_e^{'}] = w_{e}^{2} / 2 \pi \\
  \\
  v[f_t^{'}] = \pi^2 w_{t}^{2} / 48.
            \end{array} \right.
  \end{equation}

The simulations are performed for three temperatures,
$T$=$0.8$, $0.9$, and $1.0$ $\epsilon / k_{\rm B}$.
This range of temperatures is selected because at
lower temperatures the interfacial width is comparable to
the average interparticle distance, and therefore
is difficult to measure accurately.
The upper bound is set by $T_c \approx 1.085 \epsilon / k_{\rm B}$.
For each value of $T$, the profiles are fit to both
$f_x$'s described above.
Near the interface, the fitting functions can hardly be distinguished.
In fact, some studies have used an error function for theoretical derivations
while
using a hyperbolic tangent function to fit their data \cite{wern97,heer98}.

We fit our interfacial profiles for data near the interface {\it and} data
deep into the bulk liquid and vapor regions.
There is no a priori requirement that a liquid/vapor density profile must be
symmetric about the center of the interface.
However, we detected no significant amounts of asymmetry.
For each temperature and system size, the simulations are run until
the interfacial profiles show a constant $\Delta^2$.

\begin{figure}[thb]
\centerline{\psfig{file=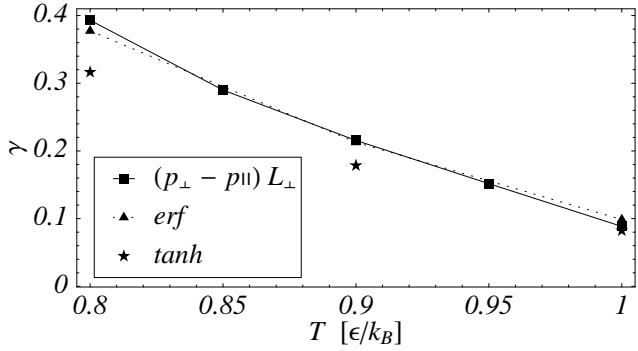,angle=270,width=\hsize,%
bbllx=150pt,bblly=75pt,bburx=510pt,bbury=740pt}}
\caption
{\label{fig_gamvsT}
Surface tension vs temperature $T$.
Results for the surface tension $\gamma_w$
obtained from fitting interfacial profiles
with a $\tanh$ or erf are compared to the values calculated from
the components of the pressure tensor,
$\gamma_p$=$L_{\perp}(p_{\perp} - p_{\parallel})$.
The error function results are in excellent agreement with $\gamma_p$
while the $\tanh$ results are systematically $15 \%$ too low.
The $\gamma_p$ value for $T$=$0.85$ is taken from
Ref.~\protect\cite{adams91}.
The $\gamma_p$ value for $T$=$0.95$ is calculated from a single
simulation with $L$=$41.9 \sigma$.
}
\end{figure}

\begin{figure}[thb]
\centerline{\psfig{file=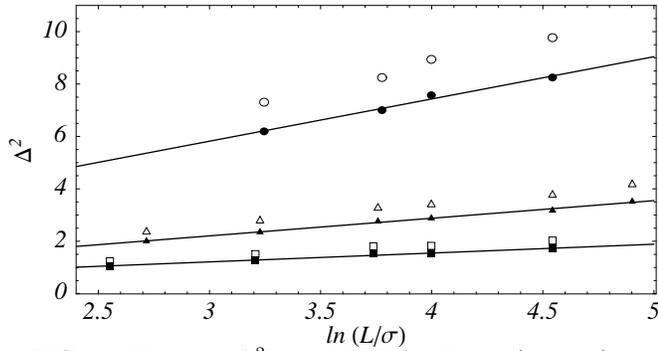,angle=270,width=\hsize,%
bbllx=135pt,bblly=35pt,bburx=520pt,bbury=772pt}}
\caption
{\label{fig_del2vsL}
Variance $\Delta^2$ versus $\ln L$ for $T$=$0.8$ (squares), $0.9$ 
(triangles) and
$1.0$ $\epsilon / k_{\rm B}$ (circles).
The open and solid symbols are obtained using hyperbolic tangent 
and error function fits to the interfacial profiles, respectively.
Lines are linear least-squares fits to the error function data.
}
\end{figure}

Figure \ref{fig_del2vsL} summarizes the analysis from our
extensive molecular dynamics simulations of a liquid/vapor interface.
For both $\tanh$ and erf fits, the data confirm a logarithmic
dependence of $\Delta^2$ on system size.
Using Eq.~(\ref{eq_delta2}) the surface tensions can be calculated from
the slopes of the best fit lines in Fig.~\ref{fig_del2vsL}
[$\gamma_w$=$k_{\rm B} T / (2 \pi \ {\rm slope})$].
The temperature dependence of the interfacial surface tensions
calculated from our simulations are shown in Fig.~\ref{fig_gamvsT}.
We compare these values of the surface tension with an independent
measurement obtained from the components of the pressure tensor,
$\gamma_p$=$L_{\perp}(p_{\perp} - p_{\parallel})$ \cite{hill86_book},
represented by solid squares.
The agreement between $\gamma_p$ and $\gamma_w$ obtained from the
error function fits is very good.
Using the $\tanh$ fits we obtain surface tensions that are systematically
$15 \%$ lower than those from the error function
fits, which follows from their larger slopes shown in Fig.~\ref{fig_del2vsL}.
Thus we obtain the somewhat unsettling result that
the value of the $\Delta^2$ and hence $\gamma_w$ depends
on the form of the fitting function used to fit $\psi(z)$.

\begin{figure}[thb]
\centerline{\psfig{file=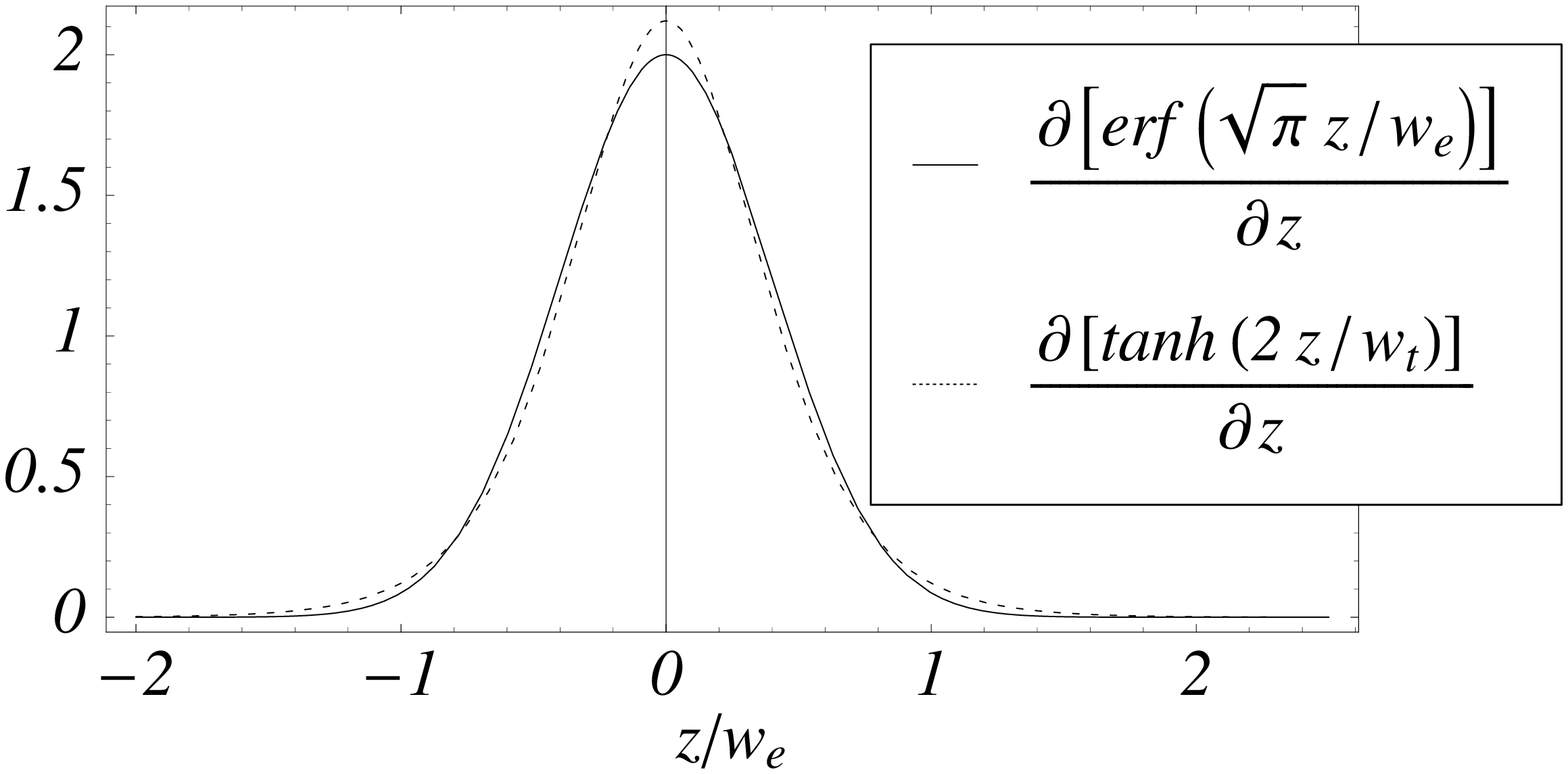,angle=270,width=\hsize,%
bbllx=150pt,bblly=80pt,bburx=500pt,bbury=760pt}}
\centerline{\psfig{file=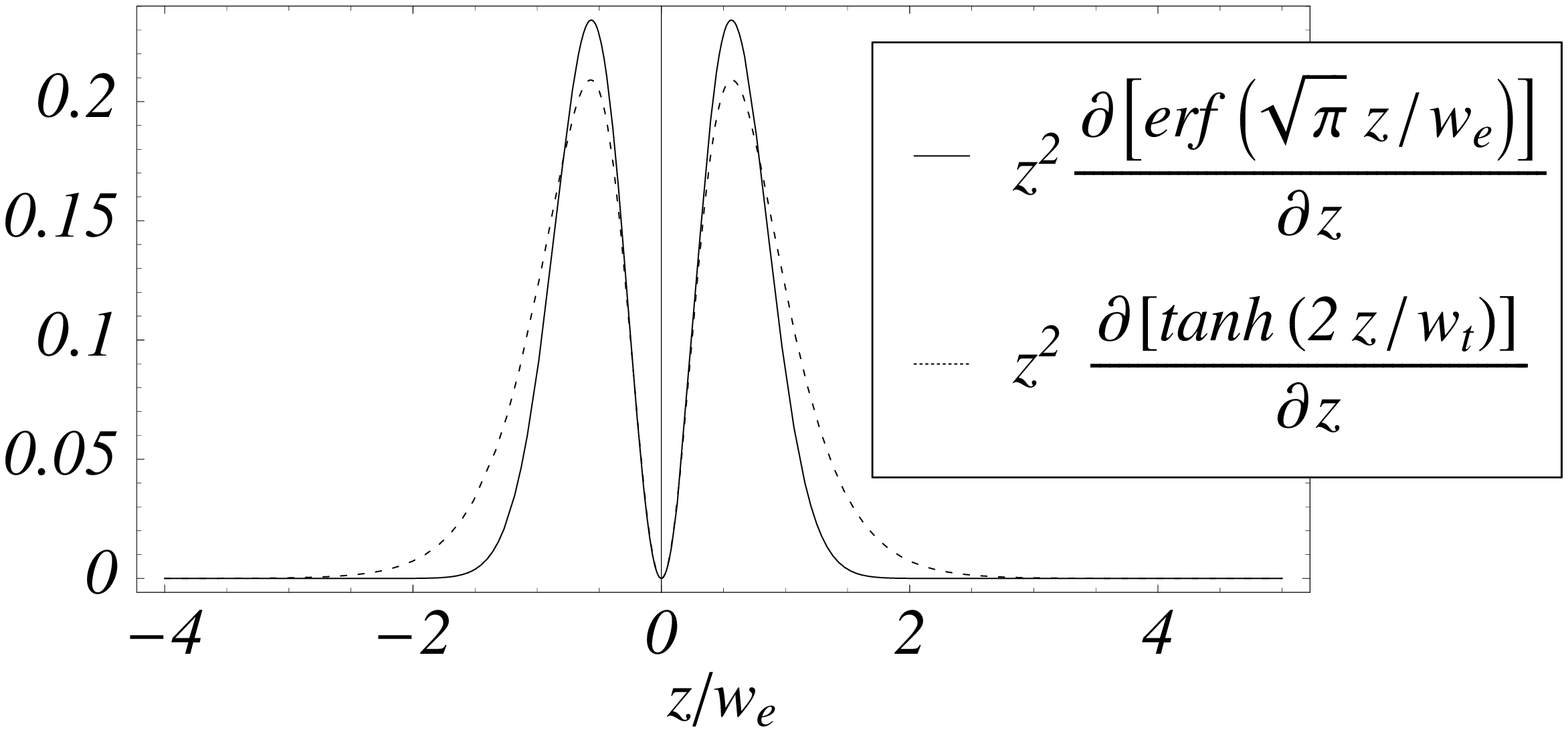,angle=270,width=\hsize,%
bbllx=165pt,bblly=75pt,bburx=490pt,bbury=760pt}}
\caption
{\label{fig_te}
Results for an unweighted fit of a $\tanh$ function
to a error function for $w_e$=$1$.
}
\end{figure}

To investigate the systematic discrepancy between
the $\tanh$ and erf fits,
we performed an unweighted fit a hyperbolic tangent function to data generated
with an error function.
The results are shown in Fig.~\ref{fig_te}.
The integrand in the numerator of the variance is plotted vs $z$
for both functions.
The integral of each of these functions is proportional to $\Delta^2$.
From Fig.~\ref{fig_te} one can see that the tails in the integrand of the
$\tanh$ function contribute more significantly than the error function,
hence the larger measured values of $\Delta^2$ from the $\tanh$ fits.
We conclude that the tails of interfacial profiles are better
captured by fits to an error function.

In this paper, we presented results of extensive molecular dynamics
simulations of liquid/vapor interfaces.
Our data confirm the capillary-wave description of the interface structure
between a Lennard-Jones liquid and its vapor phase.
When measuring the interfacial width by using second moments
of the interfacial profile derivatives, we can extract
values for the surface tension that agree very well with
calculations from the components of the pressure tensor.
We have also shown that the more robust method of extracting
the second moment is through fits to an error function, since
using a hyperbolic tangent leads to systematic errors.

The results presented here are for an isolated liquid/vapor interface.
The width of the liquid and vapor regions were
carefully chosen so that there was no interference between
the two interfaces.
An interesting extension of this work
is to study the effect of a nearby substrate on an interface.
The effect of a wall on the interface
can be modeled by adding a potential
energy term to the interface Hamiltonian, ${\mathcal{H}}[d]$,
which depends on the distance $d$ between the interface and
the wall.
This term is calculated by integrating the potential energy
between the microscopic constituents of two macroscopic objects, i.e.
the interface and a semi-infinite wall.
For pure LJ interactions, the potential energy is proportional to
$A/d^2$, where $A$ is the Hamaker constant \cite{israe91}.
The Hamaker constant contains information about the strength
of the microscopic potential, geometrical factors, and macroscopic
properties of the wall.
Since this additional term in the
Hamiltonian is quadratic in $d$, ${\mathcal{H}}$ can therefore be
diagonalized by a Fourier transform and the derivation of the capillary-wave
spectrum is similar to the one presented here.
The effect of the substrate is to cut off the long wavelength 
capillary-wave fluctuations
so that $\Delta^2$ no longer depends on $L_\parallel$
for small $d$.
The interplay between $L_\parallel$ and $d$
is an interesting question for which computer simulations
such as these can directly address.

We thank Frank van Swol for helpful discussions.
Sandia is a multiprogram laboratory operated by Sandia Corporation, a Lockheed 
Martin Company, for the United States Department of Energy under Contract 
DE-AC04-94AL85000.

\bibliographystyle{/usr/local/teTeX/texmf/tex/latex/RevTeX/prsty}

\begin{thebibliography}{10}
\bibitem{gravity}
For systems in a gravitational field $g$, a gravitational length defined by
  $l_{g}^2$=$2 \gamma / g \Delta \rho$ determines the upper cutoff, where
  $\Delta \rho$ is the difference in bulk density between the two phases.
\end{thebibliography}

\end{document}